\documentclass[conference]{IEEEtran}
\usepackage{amsmath}
\IEEEoverridecommandlockouts
\usepackage{cite}
\usepackage{amsmath,amssymb,amsfonts}
\usepackage{algorithmic}
\usepackage{graphicx}
\usepackage{textcomp}
\usepackage{xcolor}
\usepackage{subfig}
\usepackage{graphicx}
\usepackage{subcaption}

\def\minwrt[#1]{\underset{#1}{\mathrm{minimize }}}
\def\argminwrt[#1]{\underset{#1}{\text{arg min }}}

\def\BibTeX{{\rm B\kern-.05em{\sc i\kern-.025em b}\kern-.08em
    T\kern-.1667em\lower.7ex\hbox{E}\kern-.125emX}}
\begin{document}

\title{All-pole centroids in the Wasserstein metric with applications to clustering of spectral densities}

\author{
	\IEEEauthorblockN{Rumeshika Pallewela and Filip Elvander \\
	}
	\IEEEauthorblockA{ Dept. of Information and Communications Engineering, Aalto University, Finland \\
	Email: firstname.lastname@aalto.fi}\thanks{This research was supported in part by the Research Council of Finland
(decision number 362787).}

}

\maketitle

\begin{abstract}
%
%
In this work, we propose a method for computing centroids, or barycenters, in the spectral Wasserstein-2 metric for sets of power spectral densities, where the barycenters are restricted to belong to the set of all-pole spectra with a certain model order. This may be interpreted as finding an autoregressive representative for sets of second-order stationary Gaussian processes. While Wasserstein, or optimal transport, barycenters have been successfully used earlier in problems of spectral estimation and clustering, the resulting barycenters are non-parametric and the complexity of representing and storing them depends on, e.g., the choice of discretization grid. In contrast, the herein proposed method yields compact, low-dimensional, and interpretable spectral centroids that can be used in downstream tasks. Computing the all-pole centroids corresponds to solving a non-convex optimization problem in the model parameters, and we present a gradient descent scheme for addressing this. Although convergence to a globally optimal point cannot be guaranteed, the sub-optimality of the obtained centroids can be quantified. The proposed method is illustrated on a problem of phoneme classification.
\end{abstract}


\section{Introduction}
Using power spectral densities (PSDs) or related frequency-domain representation as features are extensively used in signal processing applications, finding use in clustering \cite{11345326}, classification \cite{Rohanian2026-dt}, source separation\cite{app16020572} and system identification\cite{11345101}.
%
For clustering and classification, there are two necessities; availability of a distance measure to compare spectra and a method of averaging.
Using spectral features as an 'average' or centroid  in classification and clustering has gained its popularity in the past couple of decades.

Arithmetic mean\cite{8060999} and geometrical mean \cite{4380582} are considered the most conventional averaging methods. However, these approaches tend to smear or split spectral features when prominent peaks undergo small frequency shifts across samples, leading to blurred averaged spectra.
Parametric model–based averaging \cite{10.1109/TSP.2011.2178601} reduces spectral oversmoothing by representing each PSD with auto-regressive (AR) or auto-regressive moving-average (ARMA) models by averaging in the parameter domain \cite{279281}. This is more meaningful than pointwise spectral averaging because the result remains model consistent and interpretable. However, such methods still ignore the geometry of the frequency axis of spectra that differ mainly by small shifts and can yield dissimilar parameters, so peak displacements are not explicitly captured.

Methods like $L_2$ distance \cite{NIPS2006_4fa177df}, Kullback-Leibler (KL) divergence\cite{NEURIPS2023_b5b4d923}, and Itakura-Saito (IS)\cite{1163930}  divergence, are widely used distance measures to compare spectral features. The $L_2$ distance measures point-wise bin differences but ignores the geometry of the frequency axis, causing small spectral shifts to be overly penalized. KL divergence models relative energy discrepancies but is asymmetric, (i.e, since the mismatch is weighted from the reference spectrum), and sensitive to near zero values, while IS divergence is well suited for spectral shape comparison yet is also asymmetric and numerically fragile.

\begin{figure}[t]
\centering
\subfloat{%
  \includegraphics[width=0.49\columnwidth]{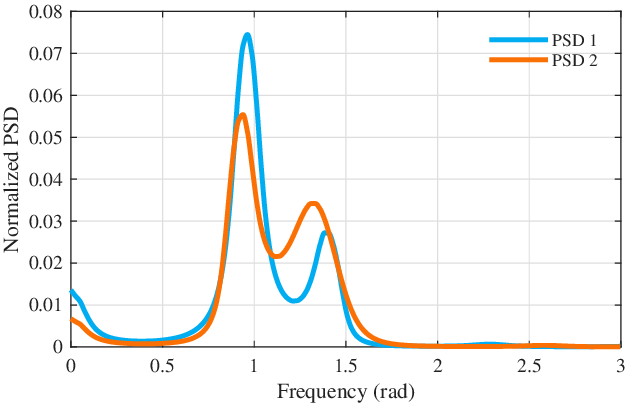}%
}\hfill
  \includegraphics[width=0.49\columnwidth]{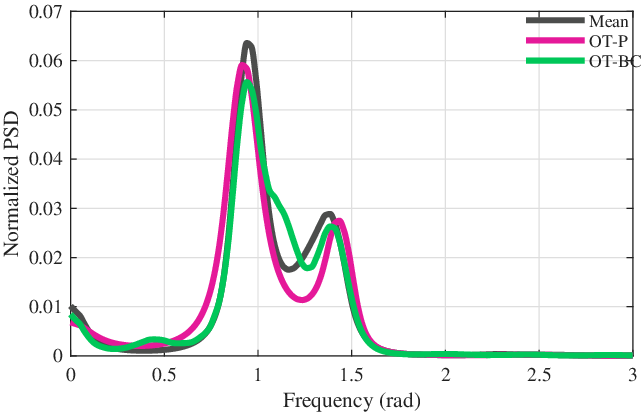}%
\caption{Phoneme /iy/: (a) two original normalized PSD samples (b) obtained using arithmetic mean, OT barycenter (OT-BC), and parametric OT barycenter (OT-P) where $P=10$.}
\label{fig:two_side_by_side_col}
\end{figure}

Optimal transport (OT) is an emerging mathematical concept that allows to compare and model non-negative mass distributions by minimizing the distance between distributions. Due to its ability to capture meaningful correspondences between distributions, OT has been successfully applied to interpolation\cite{pallewela2025roomimpulseresponseestimation}, used as a distances matrix\cite{8443147}, and also utilized for averaging spectra using OT barycenter\cite{Georgiou2009859,Gray1980367,ELVANDER2020107474} while accounting for the spectral shifts in recent literature related to signal processing and machine learning\cite{7974883}. 

Even though OT based methods can produce a meaningful “average” (centroid), the centroid is a single representation of an entire set of spectra and can therefore be complex to interpret and analyze further. In our work, we use AR models, due to their application in a wide range of areas such as in speech processing, spectrum analysis and, room acoustics\cite{10890587,NEURIPS2023_6034a661}, to obtain a low-dimensional, order capped OT barycenter which can be used in clustering and summarizing information with the use of its inherited AR model properties. Order constraining of rational spectra has been studied in \cite{carlson2020newmetricsrationalspectra}, but transport is performed  directly between pole locations, whereas we operate in the spectral domain and subsequently constrain the resulting representative to an AR model of order $p$. This yields centroids with an explicit model order, improved interpretability, and compatibility with standard signal processing tools. The optimization problem at hand is non-convex, therefore we utilize gradient descent with multi-initialization techniques in order to calculate the centroid, which is further differentiable. 

\section{Signal model}\label{sigmod}
Let $\mathcal{X} = \{x_k \mid k = 1,\ldots,K \}$ be a collection of $K$ Gaussian, zero-mean, wide-sense stationary (WSS) stochastic processes, or signals.
%
Each process is characterized  by its power spectrum $\Phi_k \in \mathcal{M}_+(\mathbb{T})$, i.e., a non-negative measure on $\mathbb{T} \triangleq [-\pi,\pi)$. Furthermore, we assume that each $\Phi_k$ is a density. Then, in order to summarize this set of stochastic processes elements, we would like to find a single representative PSD, or centroid, for the set $\mathcal{X}$. Given a distance measure $D: \mathcal{M}_+(\mathbb{T}) \times \mathcal{M}_+(\mathbb{T}) \to \mathbb{R}$, this can be defined as an element $\Phi_0 \in \mathcal{M}_+(\mathbb{T})$ minimizing the average distance to the set of PSDs $\{ \Phi_k \}_{k=1}^K$. That is,
\begin{align}\label{eq:centroid}
    \Phi_0 = \argminwrt[\Phi\in\mathcal{M}_+(\mathbb{T})] \quad \sum_{k =1}^K \frac{1}{K}D(\Phi,\Phi_k)
\end{align}
%
%
%
%
The formulation \eqref{eq:centroid} can be readily extended to weighted averages, but we restrict attention to the uniform case for simplicity.
Common choices for the distance $D$ include the $L_2$ distance, KL divergence, as well as the IS distance. The $L_2$ distance is a standard baseline for spectral matching and clustering \cite{9825708}, 
KL is widely used for distribution comparison in information theoretic learning \cite{11363202}, and IS is a classical choice in speech and audio spectral modeling and source separation due to its scale invariant emphasis on spectral shape \cite{DiffusionFrame}.

%
However, $L_2$, KL, and IS penalize amplitude differences at matching frequencies, and as a result, small peak or formant shifts that occur across speakers can be interpreted as large mismatches, producing centroids that \emph{smear} or \emph{split} spectral features. By contrast, OT treats a PSD as a distribution over frequency and permits energy to be \emph{moved} between nearby frequencies under a ground cost, making it robust to such misalignments and yielding barycenters that preserve sharp structure by effectively \emph{shifting} peaks rather than averaging them in place. 
\subsection{Optimal transport}

Optimal transport provides a principled framework for comparing
nonnegative \emph{mass distributions} by determining the least costly way to
transform one distribution into another \cite{villani2009optimal,peyré2020computationaloptimaltransport}. The resulting minimum transportation
cost defines a distance between distributions and induces a geometric structure
on the underlying domain \cite{villani2009optimal}. These ideas have also been extended as to endow the space of power spectra with a metric structure \cite{Georgiou2009859}. This allows us to define an OT distance, $D_{\mathrm{OT}}$, between two PSDs $\Phi_1$ and $\Phi_2$ by means of the Monge-Kantorovich problem
\begin{equation*}
\begin{aligned}
    D_{\mathrm{OT}}(\Phi_1,\Phi_2) = \min_{m \in \Omega(\Phi_1,\Phi_2)} \int_{\mathbb{T}\times \mathbb{T}} c(\omega_1,\omega_2) m(\omega_1,\omega_2)d\omega_1 d\omega_2,
\end{aligned}
\end{equation*}
where $c:\mathbb{T}\times \mathbb{T} \to \mathbb{R}$ is the so-called ground-cost, and where
\begin{align*}
    &\Omega(\Phi_1,\Phi_2)
    \\&= \left\{ m \in \mathcal{M}_+(\mathbb{T}\times \mathbb{T}) \mid \int_\mathbb{T} m(\cdot,\omega_2) = \Phi_1 \,\; \int_\mathbb{T} m(\omega_1,\cdot) = \Phi_2 \right\}.
\end{align*}
Here, $m$, which is a distribution on the product space $\mathbb{T}\times\mathbb{T}$, is referred to as a transport plan, and describes how mass is moved between $\Phi_1$ and $\Phi_2$. The set $\Omega(\Phi_1,\Phi_2)$ contains all valid transport plans between $\Phi_1$ and $\Phi_2$, i.e., that make sure that all mass is accounted for. As the cost of moving a single unit mass is given by the ground-cost $c$, while $D_{\mathrm{OT}}(\Phi_1,\Phi_2)$ corresponds to the most cost efficient way of morphing $\Phi_1$ into $\Phi_2$. In this work, we pick $c(\omega_1,\omega_2) = | (\omega_1-\omega_2)_{\mathrm{mod }2\pi}|^2$, i.e., the ground-cost is the squared distance on the circle. With this choice, $D_{\mathrm{OT}}(\Phi_1,\Phi_2)^{1/2}$ is a metric on $\mathcal{M}_+(\mathbb{T})$, which can be identified with the Wasserstein-2 distance \cite{Georgiou2009859}.

With this, one may define OT centroids, or barycenters, by using $D = D_{\mathrm{OT}}$ in \eqref{eq:centroid}. It may be noted that this is a linear program, albeit an infinite-dimensional one as the optimization variables are elements of $\mathcal{M}_+(\mathbb{T})$ and $\mathcal{M}_+(\mathbb{T}\times \mathbb{T})$. This type of barycenter has been considered in \cite{ELVANDER2020107474,Cazelles2019TheWD}. However, in this work, we add the additional restriction that $\Phi_0$ must correspond to an $\mathrm{AR}(P)$-model, for a given model order $P$. That is, we require that
\begin{align*}
    \Phi_0(\omega) = \frac{\sigma^2}{\left| A(e^{j\omega}) \right|^2}
\end{align*}
for some $\sigma^2 > 0$ and trigonometric polynomial $A(z) = 1+\sum_{p=1}^Pa_p z^{-p}$, where $a_p \in \mathbb{R}$, $p = 1,\ldots,P$. Consequently, we also require that the model is stable, i.e., all poles are strictly inside the unit disk, for the PSD to exist. This constraint is very useful due to its ability to produce a compact, low dimensional and interpretable representation of the spectrum. Furthermore, the all-pole model encodes dominant resonances via pole locations, and its parameters can be estimated efficiently and robustly from short data records using classical parametric spectral estimation (e.g., Yule--Walker/Burg). Such AR spectral models underpin linear predictive coding and are widely used in speech/audio processing for tasks such as speech enhancement and parametric spectrum estimation in general\cite{Stoica2005SpectralAO,10890587}.

Although this condition allows for representing $\Phi_0$ by $P+1$ real-valued parameters, computing $\Phi_0$ is no longer a convex problem as $D_{\mathrm{OT}}$ is not convex in the parameters $\left( \sigma^2, \{ a_p\}_{p=1}^P \right)$. In the next section, we propose to address this by performing gradient descent on a discretized and entropy-regularized version of \ref{eq:centroid}. Furthermore, we show that one can pick a reasonable initial point for the algorithm by means of convex techniques, as well as bound the sub-optimality of any limit point of the gradient descent scheme.

\section{Proposed method}\label{sec:method}

\subsection{Discretization and entropic OT barycenter}\label{subsec:disc_bary}


We discretize the cost function in order to obtain a differentiable function for convex programming.
Let $\mathbb{T}$ be discretized into a uniform grid of $N$ frequency points.
Each PSD is represented as a nonnegative vector on this grid,
\begin{equation}
\boldsymbol{\Phi}_0\in \mathbb{R}_+^N,
\qquad
\boldsymbol{\Phi}_k \in \mathbb{R}_+^N.
\end{equation}
obtained by normalizing the spectrum. Here $\boldsymbol{\Phi}_k \in \mathbb{R}_+^N$ denotes the $k$-th discretized and normalized PSD, and $\boldsymbol{\Phi}_0\in \mathbb{R}_+^N,$ is the barycenter  to be estimated. Additionally, $\mathbf{C}\in\mathbb{R}^{N\times N}$ is the ground-cost matrix with elements defined as $[\mathbf{C}]_{n,\ell} = c(\omega_n,\omega_\ell)$, where $c$ is the ground-cost defined earlier, and $\omega_n,\omega_\ell$ are discretization points of $\mathbb{T}$.

To obtain a smooth objective and a scalable solver, we use the entropically regularized OT cost.
Given ${\boldsymbol{\Phi}}_0,{\boldsymbol{\Phi}}_k\in \mathbb{R}_+^N,$, define
\begin{equation}
\begin{split}
D_{\mathrm{OT}\varepsilon}(\boldsymbol{\Phi}_0,\boldsymbol{\Phi}_k)
\;&=\;
\min_{\mathbf{\Pi}\ge 0}\;
\langle \mathbf{C},\mathbf{\Pi}\rangle
+\varepsilon \sum_{n,\ell} \mathbf{\Pi}_{n\ell}\bigl(\log \mathbf{\Pi}_{n\ell}-1\bigr)\\
&\text{s.t.}\;\;
\mathbf{\Pi}\mathbf{1}=\boldsymbol{\Phi}_0,\qquad \mathbf{\Pi}^\top\mathbf{1}=\boldsymbol{\Phi}_k ,
\end{split}
\label{eq:entropic_ot_primal}
\end{equation}
where $\langle \mathbf{C},\mathbf{\Pi}\rangle=\sum_{n,\ell} [\mathbf{C}]_{n\ell}\mathbf{\Pi}_{n\ell}$ and $\varepsilon>0$ controls the
amount of smoothing.
Let $\{\boldsymbol{\Phi}_k\}_{k=1}^{K}\subset\mathbb{R}_+^N$ be the discretized PSDs. The entropic OT barycenter is defined as
\begin{equation}
\boldsymbol{\Phi}_0^\star
\;=\;
\arg\min_{\boldsymbol{\Phi_0}\in \mathbb{R}_+^N}\;\sum_{k=1}^{K}\frac{1}{K}\, D_{\mathrm{OT}\varepsilon}(\boldsymbol{\Phi_0},\boldsymbol{\Phi}_k)
\label{eq:obj}
\end{equation}
Introducing dual potentials
$\mathbf{f}_k,\mathbf{g}_k\in\mathbb{R}^N$ for each coupling, we can obtain entropic barycenter dual
\begin{equation}
\begin{aligned}
\max_{\{\mathbf{\mathbf{f}_k},\mathbf{g}_k\}_{k=1}^K}\quad
&\sum_{k=1}^K \frac{1}{K}\Bigl(
\langle \mathbf{g}_k,\; \boldsymbol{\Phi}_k^{(k)}\rangle
-\varepsilon\,\bigl\langle (\mathbf{K}_\varepsilon \, e^{\mathbf{g}_k/\varepsilon},\; e^{\mathbf{\mathbf{f}_k}/\varepsilon}\bigr\rangle
\Bigr)\\
\text{s.t.}\quad &\sum_{k=1}^K \frac{1}{K} \mathbf{\mathbf{f}_k} = 0,
\end{aligned}
\label{eq:bary_dual}
\end{equation}
where $\mathbf{K}_\varepsilon$ is the Gibbs' kernel defined as:
\begin{equation}
\mathbf{K}_\varepsilon \;\triangleq\; \exp\!\left(-\frac{C}{\varepsilon}\right),
\label{eq:Gibbs_kernel}
\end{equation}
where the exponential is taken elementwise.
\eqref{eq:bary_dual} has the same optimal value as \eqref{eq:entropic_ot_primal}.

At optimality, each coupling admits the form
\begin{equation}
\mathbf{\Pi}_k^\star \;=\; \mathrm{diag}(\mathbf{u}_k)\,(\mathbf{K}_\varepsilon)\,\mathrm{diag}(\mathbf{v}_k),
\qquad
\mathbf{u}_k,\mathbf{v}_k\in\mathbb{R}^N_{+},
\label{eq:scalings}
\end{equation}
and the dual potentials related to the scalings as
\begin{equation}
\mathbf{f}_k^\star \;=\; \varepsilon \log \mathbf{u}_k,
\qquad
\mathbf{g}_k^\star \;=\; \varepsilon \log \mathbf{v}_k,
\label{eq:potentials_from_uv}
\end{equation}
 where $\log(\cdot)$ is to be interpreted element-wise.
The shared marginal constraints imply
\begin{equation}
\boldsymbol{\Phi}_0^\star \;=\; \mathbf{\Pi}_k^\star\mathbf{1}
\;=\; \mathbf{u}_k\odot ((\mathbf{K}_\varepsilon) \mathbf{v}_k),
\qquad \forall k,
\label{eq:shared_marginal}
\end{equation}
where $\odot$ denotes elementwise multiplication.
In practice, $(\mathbf{u}_k,\mathbf{v}_k)$ are computed by Sinkhorn fixed-point iterations.
We implement these updates stably in the log domain using log-sum-exp operations\cite{peyré2020computationaloptimaltransport}. A key property of entropic OT is for fixed $\boldsymbol{\Phi}_0,\boldsymbol{\Phi}_k \in \mathbb{R}_+^N$, $D_{\mathrm{OT}\varepsilon}(\boldsymbol{\Phi}_0,\boldsymbol{\Phi}_k)$ is differentiable with respect
to its marginals. If $(\boldsymbol{f}_k^\star,\boldsymbol{g}_k^\star)$ denote an optimal pair of dual
potentials, then the gradient of
$D_{\mathrm{OT}\varepsilon}(\boldsymbol{\Phi}_0,\boldsymbol{\Phi}_k)$ with respect to 
$\boldsymbol{\Phi}_0$ is represented by the dual potential $\boldsymbol{f}_k^\star$,
which is defined up to an additive constant.
Since $\boldsymbol{\Phi}_0$ satisfies the simplex constraint $\mathbf{1}^\top \boldsymbol{\Phi}_0=1$, any small update must preserve this sum; hence the update direction $b$ must satisfy
$\sum_{i=1}^N b_i = 0$ as required for first order feasibility in projected or mirror descent methods 
over the simplex \cite{BECK2003167}.
We therefore fix a unique representative by enforcing a zero-sum convention,
\begin{equation}
\tilde{\boldsymbol{f}}_k^\star
=
\boldsymbol{f}_k^\star -
\frac{1}{N}\sum_{i=1}^N f_{k,i}^\star,
\qquad
\sum_{i=1}^N \tilde f_{k,i}^\star = 0 .
\label{eq:center_f_fixed}
\end{equation}

We define the final gradient of the barycenter objective as
\begin{equation}
%
\nabla_{\boldsymbol{\Phi_0}} \left( \sum_{k=1}^K \frac{1}{K} D_{\mathrm{OT}\varepsilon}(\boldsymbol{\Phi}_0,\boldsymbol{\Phi}_k) \right) = \frac{1}{K}\sum_{k=1}^K \tilde{\boldsymbol{f}}_k^\star.
\label{eq:final_grad}
\end{equation}
which is used as the descent direction in the optimization.

\subsection{Algorithm}\label{subsec:algorithm}

Directly optimizing over AR coefficients is numerically delicate because stability constraints are nonlinear. We therefore optimize over an unconstrained parameter vector $\boldsymbol{\theta} = [\theta_0, \theta_1, \dots, \theta_P]^\top \in \mathbb{R}^{P+1}$, which represent the AR parameters $(\mathbf a,\sigma^2)$ as follows. This vector is partitioned such that $\theta_0$ determines the model gain, while the remaining components $[\theta_1, \dots, \theta_P]^\top$ define the spectral shape via reflection (PARCOR) coefficients. The corresponding AR coefficients are obtained from an intermediate parametrization, $\boldsymbol{\kappa}$ via the Levinson--Durbin recursion, such that 
coefficients are set to $\boldsymbol{\kappa_p}=\tanh(\boldsymbol{\theta_p})$, where $\tanh(\cdot)$ is applied elementwise, so that $|\kappa_p|<1$ for $p=1,\ldots,P$,
yielding a stable AR polynomial.

The parameter $\sigma^2$ (equivalently $\theta_0$) is then set so that the total integral of the PSD $\boldsymbol{\Phi}_{\boldsymbol{\theta}}$ is equal to that of the set $\{ \boldsymbol{\Phi}_k \}_{k=1}^K$.


As a deterministic initializer (denoted by $(0)$), we compute the autocovariance sequence
up to lag $P$ (from the inverse DFT of the nonparametric barycenter PSD) and solve the
Yule--Walker equations to obtain $(\mathbf a^{(0)},(\sigma^2)^{(0)})$. We then convert
$\mathbf a^{(0)}$ to reflection coefficients $\boldsymbol{\kappa}^{(0)}$ using the
step-down recursion, and initialize the unconstrained shape variables as
\[
\boldsymbol{\theta}_{1:P}^{(0)}=\operatorname{arctanh}\!\big(\boldsymbol{\kappa}^{(0)}\big),
\]
where $\operatorname{arctanh}(\cdot)$ is applied elementwise. 
This provides a stable initialization for the subsequent OT-based optimization.

Furthermore, we fit the parametric spectrum by minimizing an entropic OT objective to a collection of $K$ target spectra 
$\{\boldsymbol{\Phi}_k\}_{k=1}^{K}\subset \mathbb{R}_+^N$ ,
\begin{equation}
J_\varepsilon(\boldsymbol{\theta})
\;=\;
\frac{1}{K}\sum_{k=1}^{K}
D_{\mathrm{OT},\varepsilon}\!\bigl(\mathbf{\Phi}_{\boldsymbol{\theta}},\boldsymbol{\Phi}_k\bigr).
\label{eq:J_entropic}
\end{equation}

Minimizing $J_\varepsilon$ therefore yields an AR($P$) barycenter whose \emph{spectral shape}
$\mathbf{\Phi}_{\boldsymbol{\theta}}$ is close (on average) to the targets in the OT geometry on the frequency grid.
%
%
%
%
%
%
%

%
At a given $\boldsymbol{\theta}$, let $\tilde{\mathbf{h}}(\boldsymbol{\theta})$ be the gradient in \eqref{eq:final_grad} with $\boldsymbol{\Phi}_0 = \boldsymbol{\Phi}_{\boldsymbol{\theta}}$. Using the chain rule, the gradient of \eqref{eq:J_entropic} with respect to $\boldsymbol{\theta}$ is given by
\begin{equation}
\nabla_{\boldsymbol{\theta}}J_\varepsilon(\boldsymbol{\theta})
=
\bigl(\nabla_{\boldsymbol{\theta}}\mathbf{\Phi}_{\boldsymbol{\theta}}\bigr)^\top
\tilde{\mathbf{h}}(\boldsymbol{\theta}),
\label{eq:theta_grad}
\end{equation}
where $\nabla_{\boldsymbol{\theta}}\mathbf{\Phi}_{\boldsymbol{\theta}}$ denotes the Jacobian of the mapping
$\boldsymbol{\theta}\mapsto\mathbf{\Phi}_{\boldsymbol{\theta}}$
(PARCOR $\rightarrow$ AR $\rightarrow$ PSD sampling
$\rightarrow$ normalization). Here, $\widetilde{\mathbf{h}}$ is treated as fixed once the Sinkhorn solves
have converged for the current iterate, while
$\nabla_{\boldsymbol{\theta}}\mathbf{\Phi}_{\boldsymbol{\theta}}$ is obtained
by backpropagating through the differentiable AR-to-PSD construction.
\begin{figure}[t]
\centering
\subfloat{\includegraphics[width=0.45\textwidth]{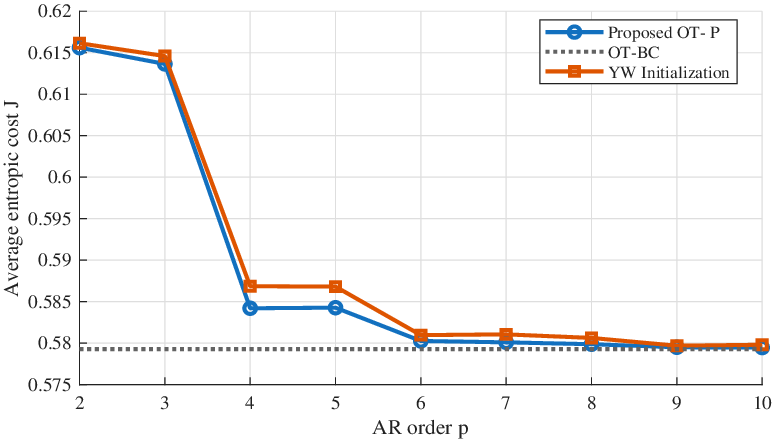}}
\caption{Comparison of average entropic costs for OT-BC, YW-initialization, and OT-P ($P=2,...,10$ and $\varepsilon =0.07$)}
\label{Numerical:ar}
\end{figure}
We update parameters using gradient descent with Armijo line search.
%
Since stability is enforced by construction through the PARCOR parametrization,
each iterate corresponds to a stable AR($P$) model, and the line search provides a robust decrease of the nonconvex
objective.


\subsection{Multi-start strategy}\label{subsec:multi}
Because the objective is nonconvex, we optimize from several different initial values 
comprising Yule–Walker initializations, perturbed Yule–Walker initializations, PARCOR-domain initializations, and Gaussian-type random stable initializations in the parameter space, and retain the best solution in terms of the final average entropic objective value.



\section{Numerical Experiments and Results}

We first illustrate the proposed approach on $4$ synthetically generated AR processes with orders incrementing from $p=10$, as shown in Fig.~\ref{Numerical:ar}. The avarage entropic cost for each methods is displayed with respect to OT-P where $p=2,...10$. The optimization converges to a suboptimal stationary point, consistent with the nonconvex formulation.

\begin{figure}[t]
\centering
\setlength{\tabcolsep}{2pt}
\renewcommand{\arraystretch}{0}

\begin{tabular}{cc}
\includegraphics[width=0.49\linewidth]{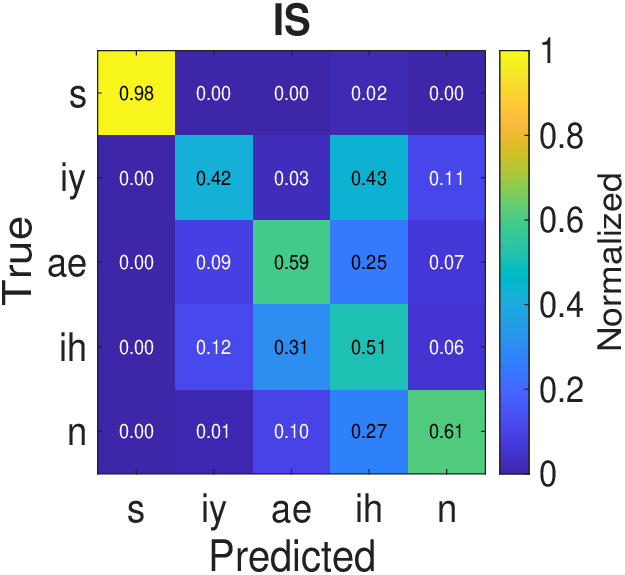} &
\includegraphics[width=0.49\linewidth]{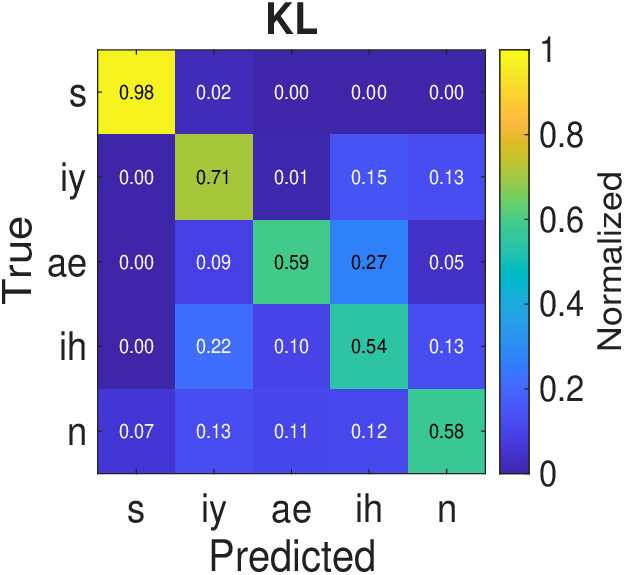} \\[+2pt]
\includegraphics[width=0.49\linewidth]{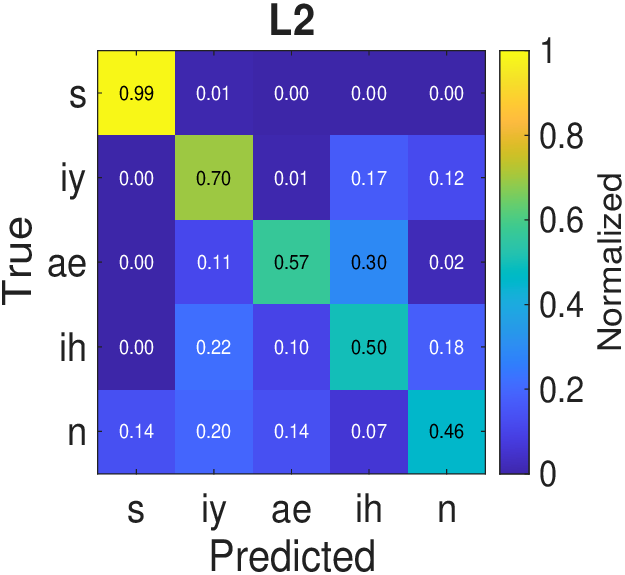} &
\includegraphics[width=0.49\linewidth]{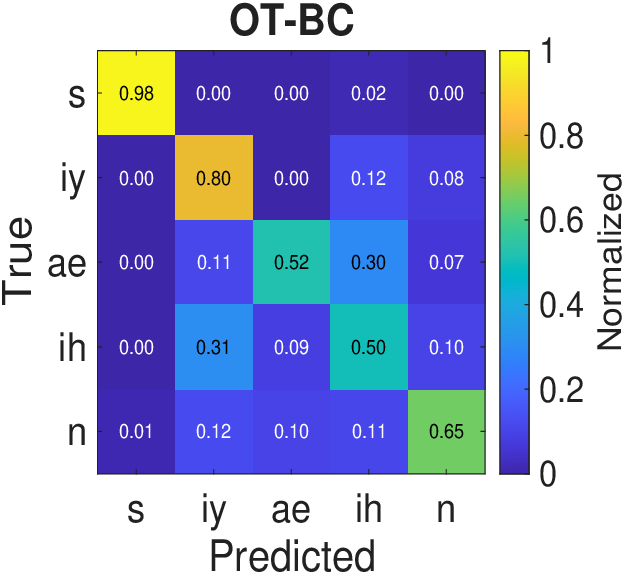} \\[+1pt]
\includegraphics[width=0.49\linewidth]{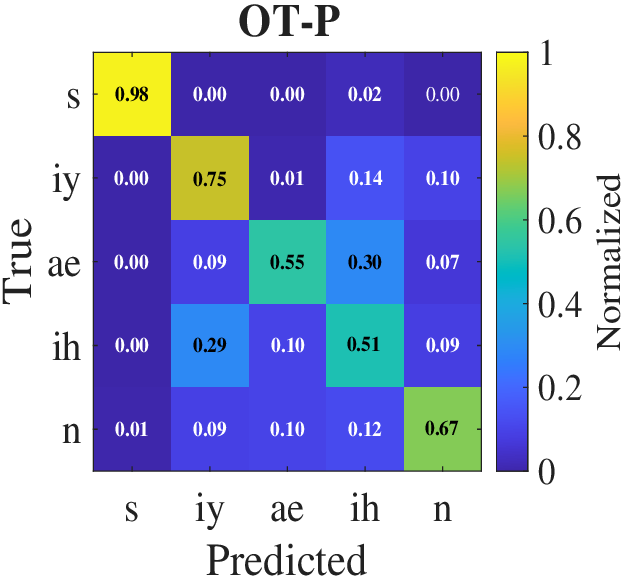} &
\phantom{\rule{0.49\linewidth}{0.49\linewidth}}
\end{tabular}

\vspace{-2mm}
\caption{Confusion matrices for multi-phoneme classification using IS, KL, $\ell_2$, OT-BC and OT-P.}
\label{Numerical:conf}
\end{figure}
We further evaluate the method on a 5-phoneme classification task (s, iy, ae, ih, n) using the TIMIT speech database \cite{TIMIT}. 
Burg-PSD features are extracted on a 128-bin frequency grid (order $10$, window $200$, hop $100$, mid-window $20$ ms), with test utterances disjoint from training data, specifically male. Since PSDs are modeled using AR$(10)$, we set the OT-P model order to $10$ for classification compatibility. 
Centroids are computed as in Sections~\ref{sigmod} and~\ref{sec:method} using $\varepsilon =0.07$ for OT-BC. Table~\ref{tab:cls_results} reports the classification results where OT-methods utilize average entropic cost as distance measure while others use their respective distances.
Performance is assessed using accuracy (ACC), balanced accuracy (BACC), F1, and AUC. While ACC measures overall correctness, BACC and F1 emphasize per-class performance, and AUC quantifies threshold-independent separability.
From Table~\ref{tab:cls_results}, OT-based methods outperform conventional spectral distances, with OT-P emerging as the most consistent. OT-P achieves the best BACC, macro-F1, and macro-AUC, while its ACC is nearly identical to OT-BC. Since ACC can be dominated by easier classes (notably /s/), improvements in BACC and F1 indicate more balanced and per-phoneme performance. Compared to KL and IS, OT-P provides clear gains in both overall correctness and class balance, confirming that incorporating frequency-axis geometry yields a more appropriate similarity measures. The confusion matrices in Fig.~\ref{Numerical:conf} clarify that all methods classify /s/ reliably, whereas the main difficulty lies in the vowels (/iy/, /ae/, /ih/) and in /n/, where spectral peaks and formants vary across speakers. IS shows strong cross-vowel confusions (especially /iy/–/ih/), and KL reduces, but does not eliminate, these errors. In contrast, OT-P exhibits a more concentrated diagonal structure across these classes, maintaining strong recall for /iy/ and improving recognition of /n/ without degrading performance on /s/. Most remaining errors occur between spectrally overlapping vowel pairs (e.g., /ae/–/ih/), reflecting their intrinsic similarity.


\begin{table}[t]
\centering
\caption{Classification performance for different spectral distance measures using test data.
}
\label{tab:cls_results}
\begin{tabular}{lcccc}
\hline
Method & ACC & BACC & F1  & AUC  \\
\hline
IS      & 0.6561 & 0.6254 & 0.6151 & 0.8503 \\
KL      & 0.7068 & 0.6790 & 0.6682 & 0.8766 \\
$\ell_2$& 0.6667 & 0.6434 & 0.6256 & 0.7726 \\
OT-BC 
        & 0.7363 & 0.6909 & 0.6859 & 0.8749 \\
OT-P 
        & 0.7342 & 0.6911 & 0.6861 & 0.8776 \\
\hline
\end{tabular}
\end{table}
\section{Conclusion}
This paper presents a geometry-aware framework for comparing and summarizing power spectral densities that integrates entropic optimal transport with low-order AR modeling. Sinkhorn barycenters are first computed and subsequently approximated by stable AR($P$) spectra using a PARCOR parametrization through minimization of an entropic OT objective. Experiments on synthetic data and a TIMIT phoneme classification task demonstrate that the proposed OT-AR method preserves spectral structure while achieving competitive performance with substantially reduced model dimensionality. Future work will consider Further optimization techniques, implementation of robust adaptive variance, extensions to multivariate and ARMA models, learned transport costs, and tighter coupling with end-to-end learning architectures.

\bibliographystyle{ieeetr}
\bibliography{ref.bib}

\end{document}